\documentclass[preprint,showkeys,showpacs,preprintnumbers,amsmath,amssymb]{revtex4}
\usepackage{graphicx}
\usepackage{graphics}
\usepackage{subfigure}
\usepackage{dcolumn}
\usepackage{bm}
\usepackage{txfonts}

\begin{document}

\preprint{}

\title{Stochastic evolutionary dynamics of minimum-effort coordination games}

\author{Kun Li$^1$}
\author{Rui Cong$^{2}$}
\author{Long Wang$^1$}
\email{longwang@pku.edu.cn} \affiliation{ $^1$Center for Systems and
Control, College of Engineering, Peking University, Beijing 100871, China\\
$^2$Department of Automation and TNList, Tsinghua University, Beijing 100084, China}

\date{\today}
\begin{abstract}
The minimum-effort coordination game, having potentially important implications in both evolutionary biology and sociology, draws recently more attention for the fact that human behavior  in this social dilemma is often inconsistent with the predictions of classic game theory. In the framework of classic game theory, any common effort level is a strict and trembling hand perfect Nash equilibrium, so that no desideratum is provided for selecting among them. Behavior experiments, however, show that the effort levels employed by subjects are inversely related to the effort costs. Here, we combine coalescence theory and evolutionary game theory to investigate this game in  finite populations. Both analytic results and individual-based simulations show that effort costs play a key role in the evolution of contribution levels, which is in good agreement with those observed experimentally. Besides well-mixed populations, set structured populations, where the population structure itself is a consequence of the evolutionary process, have  also been taken into consideration. Therein we find that large number of sets and moderate migration rate  greatly promote effort levels, especially for high effort costs. Our results may provide theoretical explanations for coordination behaviors observed in real life from an evolutionary perspective.
\end{abstract}

 \maketitle
\section{Introduction}
In game theory, coordination games are a class of games with multiple Pareto-ranked equilibria in which players choose the same or corresponding strategies~\cite{VANHUYCK1990AER,COOPER1992QJOE,Monderer1996GAEB,Jackson2002GAEB,Szolnoki2009PRE}. For those macroeconomists who believe that an economy may become mired in a low-output equilibrium, the coordination game is a particularly important paradigm~\cite{BRYANT1983QJOE,COOPER1988QJOE}. It is also worth noting that coordination game has a potentially important application to evolutionary biology~\cite{BRYANT1983QJOE,Iyer2014PRE,Iyer2014aPO}. In addition, the coordination model helps to explain the puzzle why cooperative behavior can emerge when only the fittest survive: for example, punishment, an important mechanism that restricts selfish behavior, in fact transforms social dilemmas (such as prisoner's dilemma game or public goods game) into coordination games~\cite{Chen2014NJP,Perc12NJP,Szolnoki11PRE1,Li2015PRE}.

In the minimum-effort coordination game (MECG), which captures key features of the kinds of coordination problems faced by firms, the common part of the payoff is determined by the minimum decision~\cite{VANHUYCK1990AER,Goeree1999PNAS,Goeree2001AER,Anderson2001GAEB,Weber2006AER,Devetag2005EL}. Two players choose ``effort levels'' from an available strategy set  and both have to perform a costly task to raise the joint production level. Each player's payoff is the minimum of the two efforts minus the cost of its own effort. When a player with an effort level $p_1$ encounters an opponent with $p_2$, the former gains the payoff $\mathbf P(p_1,p_2)=min(p_1,p_2)-\kappa p_1 $, where $\kappa <1$ is a cost parameter. Apparently, any common effort constitutes a Nash equilibrium, and without further refinement the Nash equilibrium concept provides little predictive power~\cite{VANHUYCK1990AER,Anderson2001GAEB,Bornstein2002GAEB,Chen2011AER}. The lowest effort level corresponds to the least efficient or risk-dominant equilibrium and the highest effort level corresponds to the Pareto-optimal or payoff-dominant equilibrium~\cite{VANHUYCK1990AER,Goeree1999PNAS,Goeree2001AER}. In the classic game theory, the set of Nash equilibria is obviously unaffected by changes in the effort cost or the number of players~\cite{VANHUYCK1990AER,Goeree1999PNAS,Anderson2001GAEB}.

However, in the MECG experiments, the observed behaviors deviate from the results predicted by game theory~\cite{Camerer1997JOEP}. In general, the cost parameter $\kappa$ plays a decisive role in the effort level: Behavior is concentrated close to the highest effort level when $\kappa$ is low, whereas high values of $\kappa$ lead to a dramatic declining of the mean contribution level. In some behavior experiments, although coordination can be improved by introducing other interaction mechanisms~\cite{Brandts2006EE,Chaudhuri2010EE,Romero2015JEBO,Hamman2007EE}, large cost continuously obstructs the stabilization of high level efforts. It is noteworthy that standard deterministic evolutionary game theory~\cite{Hofbauer98book}, as well as classic game theory, can not present explanations  for this paradoxical results either, since any common effort level is a strict Nash equilibrium and hence a stable equilibrium point for the replicator equations~\cite{MaynardSmith82book,Szabo07review}. To explain the behavior found empirically in the MECG, a variety of  theoretical approaches have been proposed. Stochastic learning models were investigated to explain the anomalous behavior observed in the MECG~\cite{Goeree2001AER,Goeree1999PNAS}. Anderson~\cite{Anderson2001GAEB} extended Rosenthal's analysis~\cite{Rosenthal89IJTG,MCKELVEY1995GAEB}  to a MECG with a continuum of actions and used the logit probabilistic choice framework to determine a unique ``logit equilibrium''. Recently, Iyer and Killingback studied the evolutionary dynamics of these games through individual-based simulations on complex networks, showing that the  evolutionary behaviors are in good agreement with those observed experimentally~\cite{Iyer2014PRE}.

 In this work, we study the MECG with mixed strategies~\cite{Tarnita2009JTB,Zhang2013SR,Szolnoki12EPL} from the perspective of the evolutionary balance between selection and mutation. Drawing lessons from coalescence theory~\cite{Antal_PNAS09,Tarnita_PNAS09}, we can well resolve the problem which effort levels are more abundant than others under any mutation rates. In the well-mixed situation, analytic results show that high effort levels can be favored when the effort cost is small, and the opposite is true when the cost is large, which is in accordance with behavior experiments and does not require extra notion of cognition or rationality. More importantly, human society is organized in forms of various sets~\cite{Tarnita_PNAS09,Fu2012SRE}, and the outcome of an evolutionary process in a structured populations (such as games on variety of networks~\cite{Nowak92Nature,Szolnoki2012SRE,Perc13review,Li2014SR}, sets~\cite{Tarnita_PNAS09,Fu2012SRE} or phenotype space~\cite{Antal_PNAS09,Fudenberg12BMB}) can be very different from that in well-mixed populations. Therefore we utilize evolutionary set theory to investigate  the extended model that individuals only interact with others who are in the same set. It seems that more sets always enhance effort levels  with all other conditions equal. This is ascribed to the fact that group structure supports the evolution of ``within-group helping''~\cite{Uyenoyama80TPB,Rogers90AN}. Meanwhile, moderate extent of random migration between sets effectively boosts mean contribution level~\cite{Tarnita_PNAS09}.

 This paper is organized as follows. Section II describes our model and methods, Sec. III presents, analyses and discusses the results, and Sec. IV draws conclusions.

\section{Model and Method}
\subsection{Basic model of the MECG}
Within this work, the basic model is a minimum-effort coordination game (MECG). Here we consider the two-person game with a strategy set $\Phi$. Without loss of generality, we set $\Phi=[0,1]$.  We specify an individual $i$'s  strategy as $p_i\in \Phi$. Let $\mathbf P(p_1, p_2)$ be the payoff that strategy $p_1$ gets from strategy $p_2$, $\mathbf P(p_1,p_2)=min(p_1,p_2)-\kappa p_1$. More specifically, $\mathbf P(p_1, p_2)$ is given by the function:
\begin{equation}
  \mathbf P(p_1,p_2)=\left\{
               \begin{array}{ll}
                 (1-\kappa)p_1, & \hbox{if $p_1\leq p_2$;} \\
                 p_2-\kappa p_1, & \hbox{if $p_1 > p_2$.}
               \end{array}
             \right.
\end{equation}
where $\kappa$ is the cost parameter.

\subsection{MECG in well-mixed populations}
At first we consider the simple but general situation of well-mixed population without any structure. In the population that consist of $N$ individuals, each one plays the MECG with every other one and they all get payoffs according to the function above. We express the fitness of player $i$ as an exponential function of the total payoff, i.e., $f_i=exp(s\mathbf P_i)$, where $s$ is the intensity of selection and $\mathbf P_i$ is $i$'s total payoff accrued through pairwise interactions with all others once. In this frequency-dependent Moran process~\cite{Nowak04Science}, at each time step an individual $i$ is selected for reproduction proportional to its $f_i$. Reproduction is subject to mutation: The offspring inherits the strategy of the parent with probability $1-u$ and with probability $u$ it adopts a strategy selected uniformly at random. A strategy $p$ is favored overall in the mutation-selection equilibrium if its abundance exceeds the mean.

In well-mixed population, we only concentrate on the situation that strategies are continuous. We use the result in Ref.~\cite{Tarnita2009JTB} to derive that, for weak selection ($s\rightarrow0$) and large population size $N$, strategy $p$ is favored by selection if $\widetilde{L}_p+Nu\widetilde{H}_p>0 $, where
\begin{eqnarray}
  \widetilde{L}_p &=& \int_0^1 \{\mathbf P(p,p)+\mathbf P(p,q)-\mathbf P(q,p)-\mathbf P(q,q)\}dq, \\
  \widetilde{H}_p &=& \int_0^1 \int_0^1 \{\mathbf P(p,r)-\mathbf P(q,r)\}dqdr.
\end{eqnarray}
 Here $\mathbf P(\cdot,\cdot)$ is defined in Eq. (1) and $p$, $q$, $r$ stand for strategy values. $\widetilde{L}_p$ and $\widetilde{H}_p$ are both functions of the unique variable $p$. Meanwhile, to determine which strategy is most favored by selection, one simply has to maximize $\widetilde{L}_p+Nu\widetilde{H}_p$. Substituting Eq. (1) into Eqs. (2) and (3), we find the condition for strategy $p$ in MECG to be favored by selection to be
\begin{equation}
  \widetilde{L}_p+Nu\widetilde{H}_p= -\frac{Nu}{2}p^2 + [(1-\kappa)Nu+ (1-2\kappa)] p+ (\frac{\kappa}{2}-\frac{1}{3}) Nu+(\kappa-\frac{1}{2})>0.
\end{equation}
We can obtain the most common strategy $p$ by maximizing $\widetilde{L}_p+Nu\widetilde{H}_p$ in inequality (4).

\subsection{MECG in set structured populations}
In this extended model, a population of finite size $N$ is distributed over $M$ sets and each individual belongs to one set. The sets could be geographical islands but they could also be phenotypic traits or tags~\cite{Tarnita11PNAS}. Two individuals interact only if they are in the same set (have the same tag). Successful sets attract more individuals. In addition, in the mutation-selection analysis, besides strategy mutation $u$, a set mutation rate $v$ is introduced to represent individuals' random migration between sets. The update rule is similar to that in well-mixed population. It is useful to consider the rescaled mutation rates, $\mu=Nu$ and $\nu=Nv$, in the theoretical analysis.

When available strategies  in the evolutionary process are discrete and the total number is $n$, for simplicity, we utilize  the $n\times n$ payoff matrix $\mathbf{A}=[a_{ij}]$ ($i,j=1,\cdots,n$) to describe payoffs between any two discrete strategies. Here, $a_{ij}=\mathbf P(p_i,p_j)$, where $\mathbf P(\cdot,\cdot)$ is defined in Eq. (1). Drawing lessons from Ref.~\cite{Tarnita11PNAS}, we acquire  that a strategy $p_m$($m=1,\cdots,n$) is favored by selection  if
\begin{equation}
  \lambda_1(a_{mm}-\overline{a_{**}})+\lambda_2(\overline{a_{m*}}-\overline{a_{*m}})+\lambda_3(\overline{a_{m*}}-\overline{a})>0,
\end{equation}
where $\overline{a_{m*}}=\sum_ja_{mj}/n$, $\overline{a_{*m}}=\sum_ja_{jm}/n$, $\overline{a_{**}} = \sum_j a_{jj}/n$, and $\bar{a} = \sum_i \sum_j a_{ij} / n^2$.
In addition, drawing supports from coalescence theory, up to the same positive constant factor,
\begin{eqnarray}
  \lambda_1 &\propto& (1+\nu)(3+\mu+\nu)(M(2+\mu)(3+3\mu+2\nu)+\nu(4+3\mu+2\nu)), \\
  \lambda_2 &\propto& M(2+\mu)(9+3\mu(4+\mu)+6\nu+5\mu\nu+\nu^2) \cr
            &+&\nu(3\mu^3+2(2+\nu)(3+\nu)^2+\mu^2(21+8\nu)+\mu(49+\nu(38+7\nu))), \\
  \lambda_3 &\propto& \mu[M(2+\mu)(9+3\mu(4+\mu)+7\nu+5\mu\nu+2\nu^2) \cr
            &+& \nu(34+3\mu^3+40\nu+2\nu^2(8+\nu)+\mu(3+\nu)(16+7\nu)+\mu^2(21+8\nu))].
\end{eqnarray}

When strategies are continuous, using the result in Ref.~\cite{Fu2012SRE} derived by Fu et al., we get that strategy $p$ is more abundant than the mean frequency if
\begin{equation}
\lambda_1\int_0^1[\mathbf P(p,p)-\mathbf P(q,q)]dq + \lambda_2\int_0^1[\mathbf P(p,q)-\mathbf P(q,p)]dq +
\lambda_3\int_0^1\int_0^1[\mathbf P(p,r)-\mathbf P(q,r)]dqdr>0.
\end{equation}
Here the $\lambda_i$ terms are the same as for the discrete strategies, and $\mathbf P(\cdot,\cdot)$, $p$, $q$, and $r$ follow the same definition as in Eq. (2) and (3).

\subsection{Individual-based simulations}
Individual-based simulations are used for the support of this weak selection analytical calculations. Meanwhile, with this tool, we can investigate  how the average effort evolves across a wide range of selection strengths. In our simulations, individuals interact in a  population of constant size $N=100$. Each individual's strategy is initialized randomly at the beginning of the simulation, and the strategies of all individuals are recorded over $10^9$ generations.

\section{Results and Analysis}

\subsection{MECG in well-mixed populations}
 In the weak selection limit where all individuals have roughly the same reproductive success, it is possible to analytically derive which strategy is most frequent. In the well-mixed populations without any structure, by maximizing $\widetilde{L}_p+Nu\widetilde{H}_p$ in condition (4), we find that the most common strategy $p_M$ follows:
\begin{equation}
 \left\{
   \begin{array}{ll}
     \kappa \leq \frac{1}{2} : \left\{
                   \begin{array}{ll}
                    p_M=1 &\hbox{(if $Nu \leq \frac{1-2\kappa}{\kappa}$)}  \\
                     p_M=\frac{(1-\kappa)Nu+(1-2\kappa)}{Nu} &\hbox{(if $Nu > \frac{1-2\kappa}{\kappa}$)}
                     \end{array}
                 \right.
\\
     \kappa>\frac{1}{2} :\left\{
             \begin{array}{ll}
               p_M=0 &\hbox{(if $Nu \leq \frac{2\kappa-1}{1-\kappa}$)} \\
              p_M=\frac{(1-\kappa)Nu+(1-2\kappa)}{Nu} &\hbox{(if $Nu > \frac{2\kappa-1}{1-\kappa}$)}.
             \end{array}
           \right.
   \end{array}
 \right.
 \end{equation}
As is shown in Fig. 1, the effort cost $\kappa$ plays the decisive role in determining  $p_M$: For small $\kappa$ ($\kappa<1/2$), natural selection most favors individuals who are willing to contribute a higher level ($p_M>1/2$) ; Whereas large $\kappa$ ($\kappa>1/2$) leads to the prevalence of low $p_M$ ($p_M<1/2$). $p_M$ monotonically increases with the decreasing of effort cost, according with economic intuition and patterns in laboratory data~\cite{VANHUYCK1990AER,COOPER1992QJOE}. For low mutation rates, at most two strategies are involved in the population. When $\kappa<1/2$ the fully contribution strategy $p=1$ behaves better than any another strategy, thereby remaining the most abundant; When $\kappa>1/2$, the risk-dominant strategy $p=0$ prevails. Conversely, for high mutation rates, all strategies have almost equal frequencies throughout evolution, and the strategy contributing $p=1-\kappa$ occupy a position of prominence in this strategy-coexistence state. Intrudingly, despite the strategy mutation rates, $p_M$ sustains at the level of 0.5 for the critical cost value  $\kappa=0.5$, resembling a `` neutral state''.

We next evaluate the impact of selection intensity $s$ on the strategy distributions in the low mutation case. Sufficiently small $u$ assures that a single mutant vanishes or fixes in a population before the next mutant appears, therefore the population is homogeneous most of the time~\cite{Fudenberg2006JOET}. Thus besides individual-based simulations, embedded Markov chain~\cite{Hauert07Science,Sigmund10Nature} can be utilized here to describe the evolutionary dynamics. Simulation results,  in good agreement with  the analytical results, show that strengthening $s$  does not alter the most common strategy $p_M$, but enhances the advantage of $p_M$. When $\kappa>1/2$  the average contribution level of the whole population decreases with the increment of $s$ (see Fig.2 (a), (b)), and the opposite is true for $\kappa<1/2$ (see Fig.2 (c), (d)). If the imitation strength is considered to be a measure of how precise people's information is in learning~\cite{Manapat12JTB,Rand13PNAS}, it seems that  information  transparency drives the majority of individuals to make a choice between the ``worst'' Nash outcome $p_M=0$ and the ``best'' equilibrium $p_M=1$, critically, relying on whether the effort cost is large.

Drawing support from individual-based simulations, we have further studied how the average effort evolves across a wide range of selection strengths and mutation rates. Some view points above have been strengthened in this more general case. As shown in Fig. 3, for small $s$, all the average strategies are close to 0.5, their neutral values. As $s$ increases, the average efforts are determined primarily by cost parameter $\kappa$. $\kappa<1/2$ leads to the fact that the mean contributions are always more than a half and ascends as  $s$ rises. On the contrary, $\kappa>1/2$ causes a detrimental effect to whole populations' efforts with increasing $s$. Strategy mutation, introducing randomness into selection, always inhibits the mean strategy from converging to the extreme states ``fully cooperation''($p=1$) or ``complete defection'' ($p=0$).

\subsection{MECG in set-structured populations}
Furthermore, we are interested in how individuals behave in the MECG if they are located in different groups and they do not play with out-group members. Here, Successful sets attract members, and individuals may  adopts a random set with probability $v$, called migration rate, similar to strategy mutation $u$. This extended model can be investigated using evolutionary set theory~\cite{Tarnita_PNAS09,Fu2012SRE}. Initially, we concentrate on the simple case of only two strategies $p_1$ and $p_2$, where $p_1>p_2$. In well-mixed populations, irrespective of any choosing intensity $s$ or mutation rate $u$, $\kappa>1/2$ determines that $p_1$ always performs worse than $p_2$. However, ``in-group favoritism'' may greatly facilitate $p_1$, if
\begin{equation}
  I \triangleq (1-\kappa)\lambda_1-\kappa\lambda_2+(\frac{1}{2}-\kappa)\lambda_3>0,
\end{equation}
which can be obtained by substituting Eq. (1) into condition (5) when the total number $n$ of available strategies is $2$ (Correspondingly the size of matrix $\mathbf A$ is $2\times 2$ ). For small strategy mutation ($\mu=Nu\rightarrow0$), $I>0$ can be further simplified as
\begin{equation}
  I_1\triangleq(1-\kappa)\lambda_1-\kappa\lambda_2>0.
\end{equation}
As shown in Fig. 4(a), for large effort cost $\kappa>\frac{1}{2}$ and high migration rate $\nu=Nv\gg\frac{1}{1-\kappa}$, condition $I_1>0$ leads to
\begin{equation}
 M>\frac{(2\kappa-1)\nu }{2(1-\kappa)},
\end{equation}
showing the minimum requirement for $M$ to ensure that $p_1$ prevails over $p_2$. As $\mu$ increases, more sets are required for sustaining $p_1$ at a higher level than $p_2$ (see Fig. 4 (b)-(d)). In this sense, for large cost parameter $\kappa$  either elevating $\mu$ or $\nu$ (when $\nu$ is large ) undermine the evolution of contribution level, but increasing $M$ always boosts effort. This can be attributed to the fact that more sets enhance the opportunity of the clustering of higher level strategy ($p_1$) to resist against the  invasion from the lower level strategy ($p_2$), even under unfavorable conditions ($\kappa>1/2$).

Moreover, in the weak selection limit, we  investigated the MECG with continuous strategy set $\Phi=[0,1]$. A global mutation
model, in which an offspring mutant adopts a strategy randomly and uniformly  drawn from the unit  interval $\Phi$, is utilized to arrive at the analytical results. Using inequality (9), We deduce that natural selection favors strategy $p$ if and only if
\begin{eqnarray}
  C(p)&\triangleq&-\frac{\lambda_3}{2}p^2+[(1-\kappa)\lambda_1-\kappa\lambda_2+(1-\kappa)\lambda_3]p \cr
  &-&\frac{1-\kappa}{2}\lambda_1 + \frac{\kappa}{2}\lambda_2+(\frac{\kappa}{2}-\frac{1}{3})\lambda_3>0.
\end{eqnarray}
Similarly, the most abundant strategy $p_M$ can be obtained by maximizing $C(p)$. Interestingly, for small strategy mutation $\mu\rightarrow0$, the fully contribution behavior ($p_M=1$) is most favored if $I_1>0$ (inequality (12)) holds; Otherwise the majority choose to invest nothing ($p_M=0$). The optimum level $p_M$ is presented in Fig. 5 across a wide range of
$\mu$, $\nu$ and $M$. Obviously, $\kappa$ also plays the key role  in determining $p_M$ (compare Fig. 5(a)-(c) with (d)-(f)). The evolution of $p_M$ is prominently promoted by increasing $M$, especially for large $\kappa$, as shown in Fig. 5(a)-(c). However, low values of set mutation rate $\nu$, which can be treated as failure in distinguishing sets, keeps individuals from adopting high effort levels (note the lower parts of Fig. 5(a)-(c)). In addition, moderate $\nu$, equipped with small $\mu$, is capable of maximizing $p_M$  even when $\kappa>1/2$ (see the left parts of Fig. 5(b)(c)). We are also interested in whether the average strategy frequency $\langle p\rangle$ of the whole population can exceed the neutral level for large $\kappa$. Drawing supports from Ref.~\cite{Fu2012SRE}, we can conclude that the condition which ensures $\langle p\rangle_{s\rightarrow0}>1/2$ is
\begin{equation}
\int_0^1pC(p)dp>0
\end{equation}
which also leads to $I>0$ (inequality (11)), coinciding with the two strategies case. The feasible regions of combinations of $M$ and $\nu$, where the whole population incline to contribute a higher level on average, are also presented in Fig. 4.

Last, under moderate selection intensity, by computer simulations we show the impacts of $M$ and migration rate $\nu$ on average effort respectively. Low $\kappa$ facilitates effort, in accordance with above analysis. For  $\kappa>1/2$, the average effort ascends sharply with $M$ when $M<100$, then stabilizes at a level slightly above $1/2$. As shown in Fig. 6(b), there is an optimal level of $\nu$ promoting the effort level. As mentioned in Ref.~\cite{Fu2012SRE}, without flexibility in group identity ($\nu\rightarrow0$) individuals reside in the same sets forever. On the other hand, excessively high values of $\nu$ make the association between strategy and group membership breaks down.

\section{Conclusions}
In the minimum-effort coordination game, any common effort constitutes a Nash equilibrium. Without further refinement, the Nash equilibrium concept provides little predictive power. However, experimental data show that efforts are much lower when effort is more costly, or when there are more players~\cite{VANHUYCK1990AER,COOPER1992QJOE,Monderer1996GAEB}. To explain the inconsistency, we bring together ideas from coalescence theory and evolutionary game dynamics, to investigate how effort levels evolve in a mutation-selection process.

In the well-mixed situation, analytical results show that effort cost is the primary factor in determining contribution levels, which is in accordance  with  experimental data. Meanwhile, changing imitation intensity can not alter the most common strategy here, different from the case in the TDG~\cite{Manapat12JTB}. When selection is weak, the dynamics depends greatly on the strategy mutation rate $u$. The most common  strategy adopted by players converges from the highest contribution level (when  $\kappa<1/2$) or the lowest contribution level (when  $\kappa>1/2$) to  $1-\kappa$ as $u$ increases.

Furthermore, populations are structured by geography or other factors~\cite{Tarnita_PNAS09}. Individuals usually can only interact with those in the same set or group. Therefore evolutionary set theory is utilized to study evolutionary dynamics of MECG in structured populations. It seems that more sets and moderate migration rate can effectively promote effort levels, especially when the costs are large. Since higher effort levels in MECG  may be considered to represent more cooperative strategies~\cite{Iyer2014PRE}, our finding strengthens the viewpoint that the structure of the population allows ``cooperative'' strategies to ``cluster'' in certain sets to resist the invasion of the ``defective'' ones. This also helps to explain why optimal coordination level is sometimes achieved even if the cost is very large in  human society.

The MECG  is of considerable practical importance since many interesting and significant real-world situations can be modeled by such games~\cite{BRYANT1983QJOE,Iyer2014PRE}. We hope our observations could help to explain how coordination behaviors evolve in a variety of real life situations.

\begin{acknowledgments}
The authors are supported by NSFC (Grants 61375120 and 61533001).
\end{acknowledgments}

\clearpage

\begin{figure}
\includegraphics[width=0.7\columnwidth]{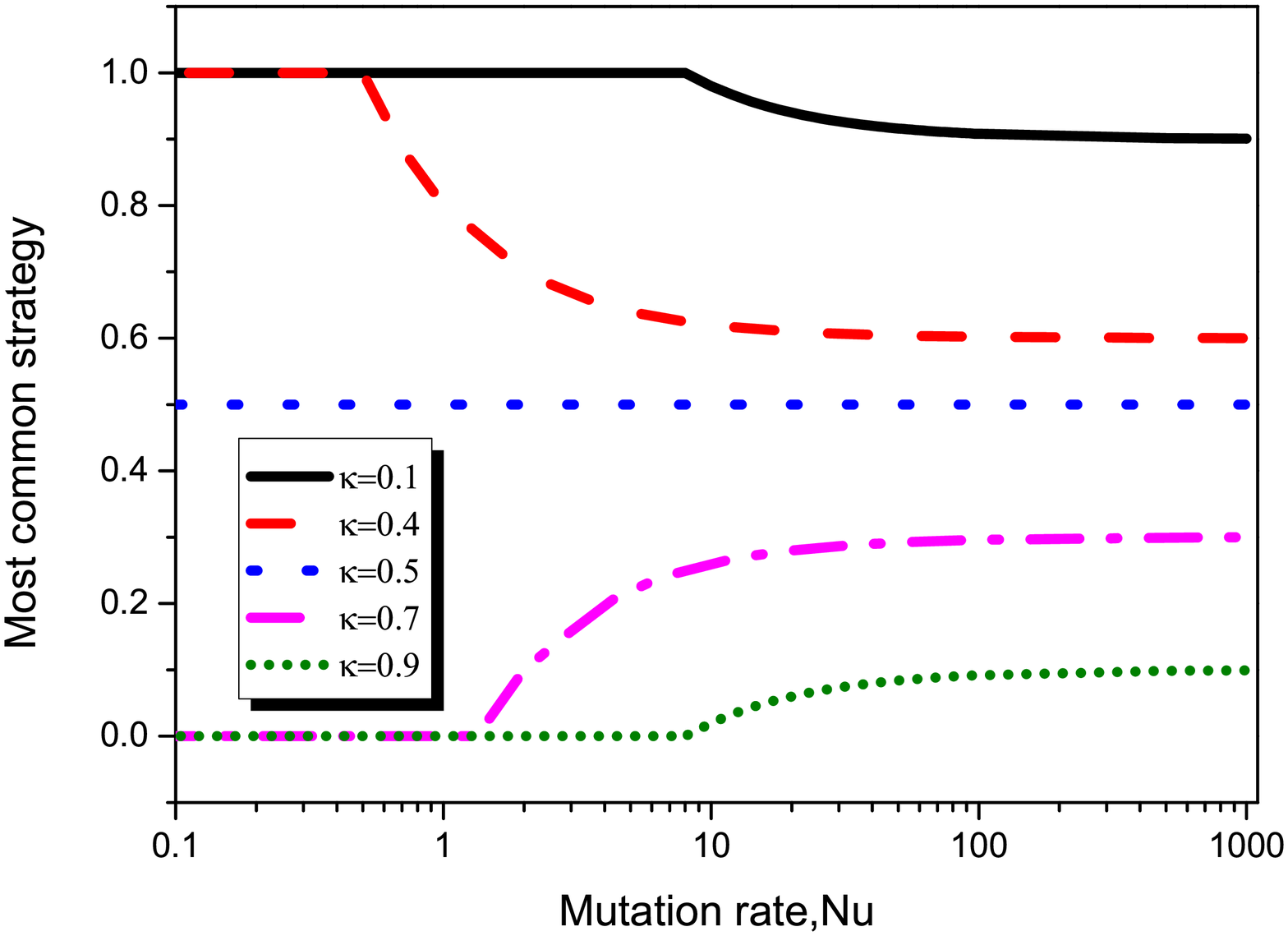}
\caption{(Color online) The most common strategy as a function of the strategy mutation rate (in terms of the expected number of mutants per generation, $Nu$) in the weak selection limit. $N$ is the total population size and $u$ is the strategy mutation rate. When the effort cost $\kappa$ is small, ``cooperative'' strategies are most common. When $\kappa$ is large, the opposite is true.
   }\label{fig1}
\end{figure}

\begin{figure}
\includegraphics[width=0.85\columnwidth]{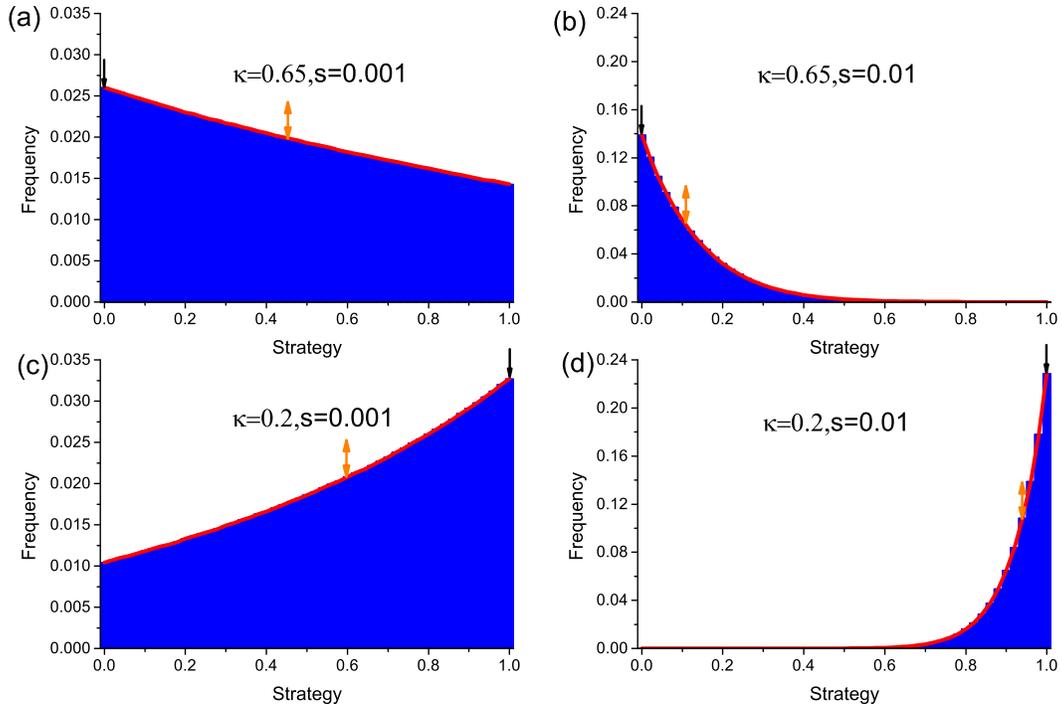}
\caption{(Color online) The strategy distributions in the low mutation case for different effort cost  $\kappa$ and selection strength $s$ values. The average strategy for each distribution is marked by a double-headed orange arrow, while the most common strategy marked by a single-headed red arrow. Results are averaged over more than $3\times10^9$ time steps. Red lines correspond to analysis results calculated by embedded Markov chain. Other parameters are $N=100$, $u=0.0001$. }\label{fig2}
\end{figure}

\begin{figure}
\includegraphics[width=0.7\columnwidth]{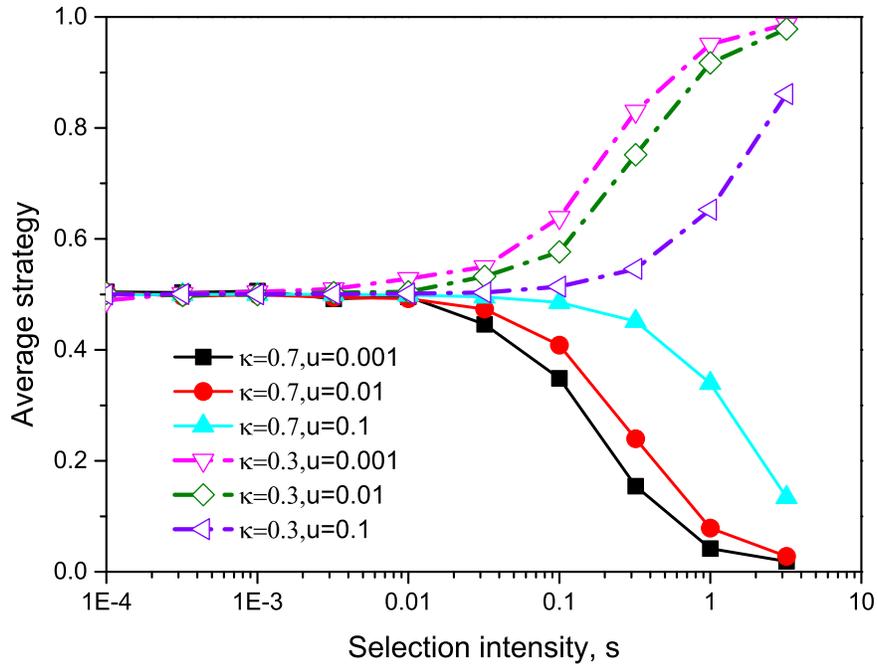}
\caption{(Color online) The average effort level as a function of the selection strength for different combinations of $\kappa$ and $u$, determined by individual-based simulations. Shown are time-averaged values over $10^9$ generations, using the population size $N=100$.}\label{fig3}
\end{figure}

\begin{figure}
\includegraphics[width=0.85\columnwidth]{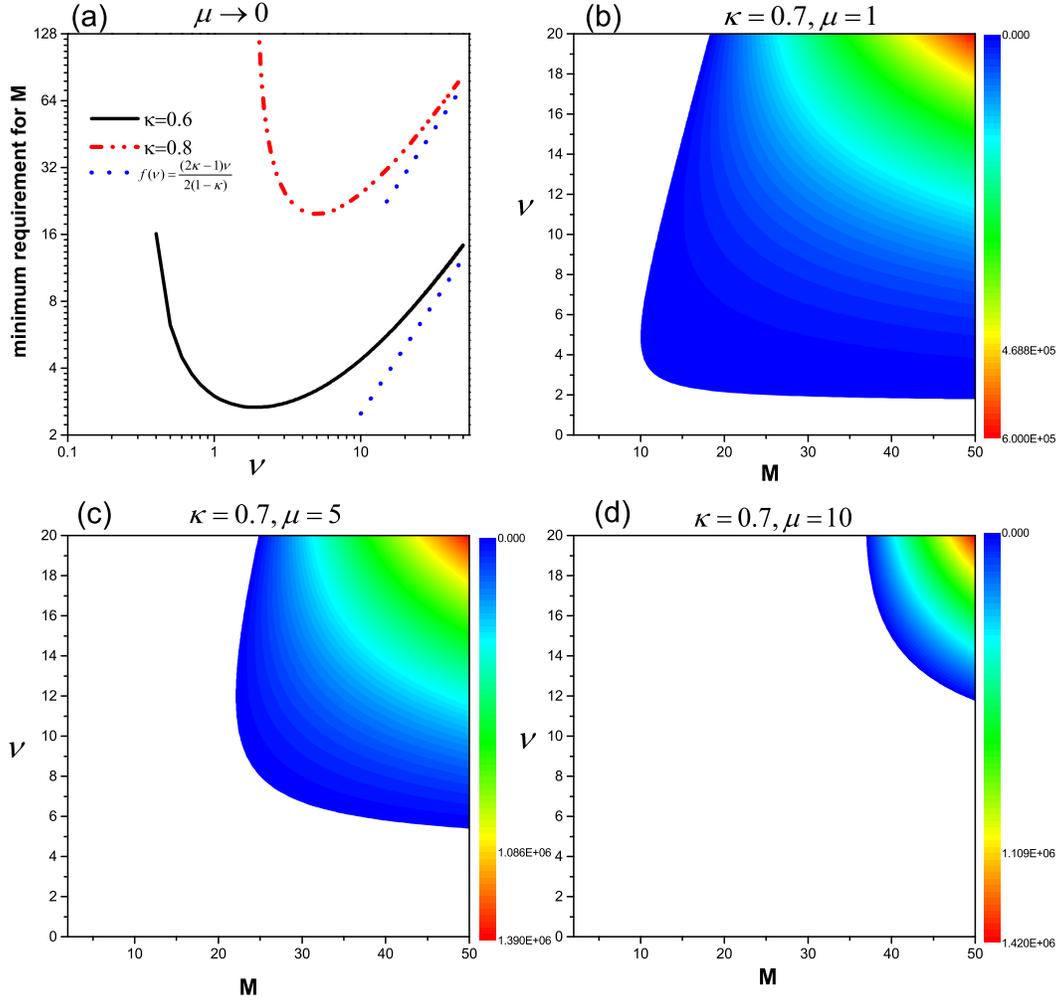}
\caption{(Color online)  (a) The minimum requirement for set number $M$ above which higher effort level is favored as a function of $\nu=Nv$. (b)-(d) The feasible regions (colorful regions) for higher effort level to prevail over lower effort level in dependence on both $\nu$ and $M$, for different combinations of $\kappa$ and $\mu=Nu$. Color bar shows exact values of $I$ in condition (11). The population size is $N=100$. }\label{fig4}
\end{figure}

\begin{figure}
\includegraphics[width=1.0\columnwidth]{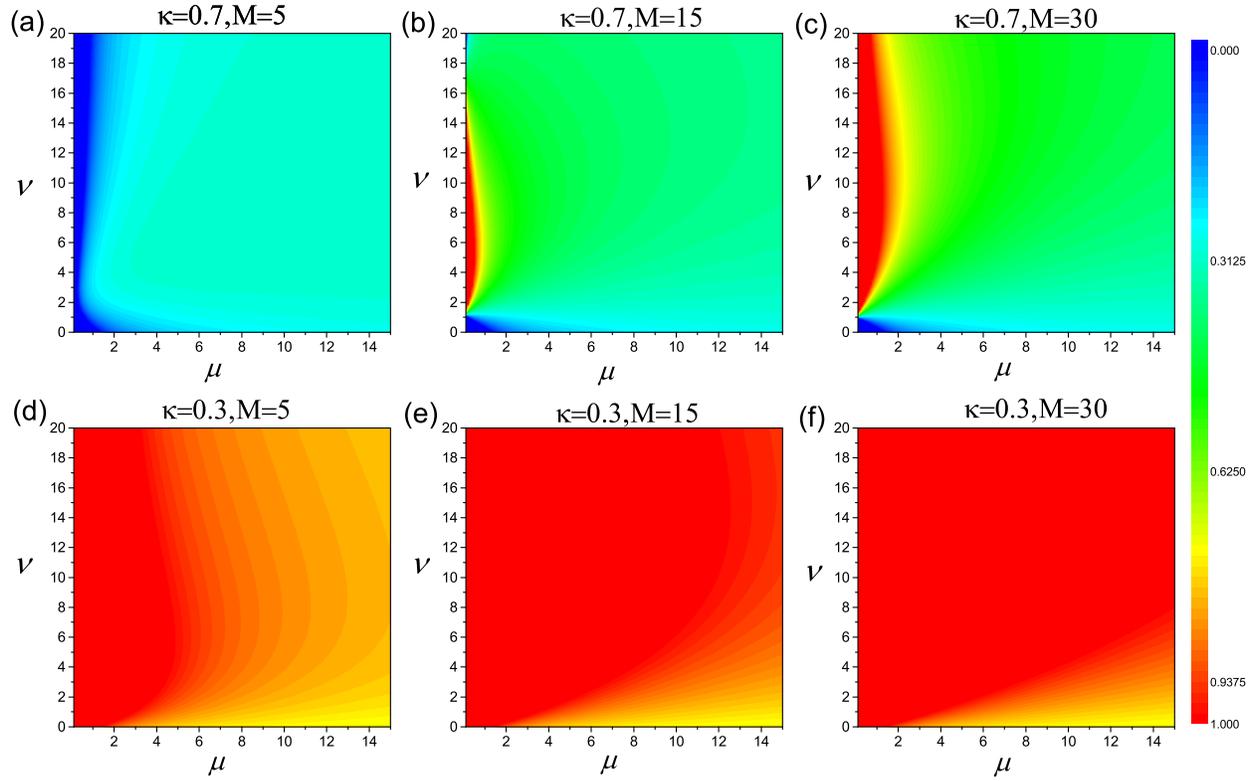}
\caption{(Color online)Influence of $\mu=Nu$ and $\nu=Nv$ on most common strategy for different values of $\kappa$ and $M$. (a)-(c)  corresponds to high effort cost parameter $\kappa=0.7$ ($\kappa>0.5$). (d)-(f)  refers to low effort cost parameter $\kappa=0.3$ ($\kappa<0.5$). Color bar shows the exact values of the most common effort levels. The population size is $N=100$.    }\label{fig5}
\end{figure}

\begin{figure}
\includegraphics[width=0.9\columnwidth]{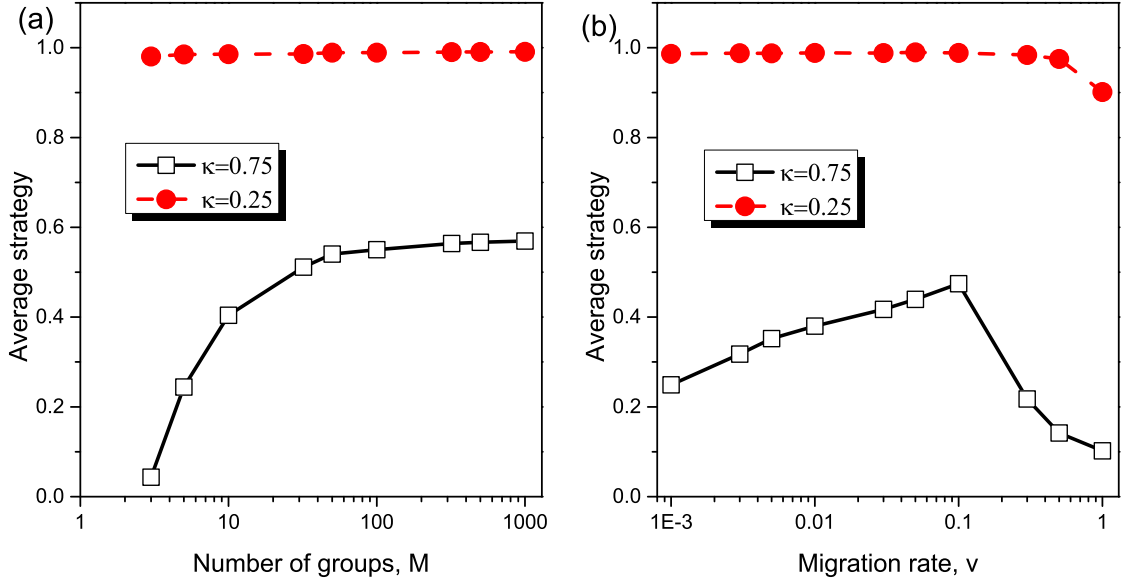}
\caption{(Color online) The average effort level as a function of (a) set number $M$ and (b) migration rate $v$, obtained by individual-based simulations. Increasing group numbers $M$ enhances ``cooperation'',especially for large $\kappa$. However, moderate migration rate promotes average effort level. Parameters: (a) $N=100$, $s=0.01$, $u=0.01$, $v=0.1$, (b) $N=100$, $s=0.01$, $u=0.01$, $M=15$. Results are averaged over $10^9$ time steps.  }\label{fig6}
\end{figure}

\end{document}